\documentclass{PoS}
\newcommand{\be}{\begin{eqnarray}}
\newcommand{\ee}{\end{eqnarray}}

\newcommand\nn{\nonumber}

\newcommand{\mat}{\left ( \begin{array}{cc}}
\newcommand{\emat}{\end{array} \right )}
\newcommand{\vect}{\left ( \begin{array}{c}}
\newcommand{\evect}{\end{array} \right )}

\newcommand{\hm}{\hat m}
\newcommand{\ha}{\hat a}
\newcommand{\hz}{\hat z}
\newcommand{\hze}{\hat \zeta}
\newcommand{\hx}{\hat x}
\newcommand{\hy}{\hat y}

\newcommand{\bmini}{\begin{minipage}}
\newcommand{\emini}{\end{minipage}}

\title{Chiral Dynamics With Wilson Fermions}

\ShortTitle{Chiral Dynamics With Wilson Fermions}

\author{\speaker{K. Splittorff}\\
        Discovery Center, Niels Bohr Institute, University of Copenhagen,
Blegdamsvej 17, DK-2100, Copenhagen {\O}, Denmark\\
        E-mail: \email{split@nbi.dk}}

\abstract{Close to the continuum the lattice spacing affects the smallest 
eigenvalues of the Wilson Dirac operator in a very specific manner
determined by the way in which the discretization breaks 
chiral symmetry. These effects can be computed analytically by means 
of Wilson chiral perturbation theory and Wilson random matrix theory. 
A number of insights on chiral Dynamics with Wilson fermions can be 
obtained from the computation of the microscopic spectrum of the 
Wilson Dirac operator. For example, the unusual volume scaling of the 
smallest eigenvalues observed in lattice simulations has a natural 
explanation. The dynamics of the eigenvalues of the Wilson Dirac 
operator also allow us to determine the additional low energy 
constants of Wilson chiral perturbation theory and to understand 
why the Sharpe-Singleton scenario is only realized in unquenched 
simulations.}

\FullConference{The 30 International Symposium on Lattice Field Theory - Lattice 2012,\\
		June 24-29, 2012\\
		Cairns, Australia}

\begin{document}

\section{Introduction}

In order to study the chiral limit in lattice QCD with Wilson 
Fermions, it is essential to understand the phase structure at nonzero 
lattice spacing. Since the continuum limit, $a\to0$, and the chiral 
limit, $m\to0$, do not commute, lattice spacing effects can induce new phase 
transitions which have no analogue in the continuum. These new phase 
structures are known as the Aoki phase \cite{Aokiclassic} and the 
Sharpe-Singleton scenario \cite{SharpeSingleton}. The presence
of new phase structures due to the lattice artifacts may at first 
appear to be a severe drawback for the application of Wilson fermions, 
however, these lattice artifacts can be turned to one's favor: The 
Aoki phase is reached through a second order phase transition and 
just outside the Aoki phase the pions have dispersion relations 
which resemble those of almost massless quarks in the continuum.
 
Because the Wilson term breaks chiral symmetry the effects of the 
lattice spacing lead to new terms in chiral perturbation theory. 
This extended low energy theory is know as Wilson chiral perturbation 
theory. The precise form of the terms was worked out in 
\cite{SharpeSingleton,RS,BRS} and at order $a^2$ involves three new 
low energy constants, which encode the severity 
of chiral symmetry breaking by the lattice artifacts. 
Recently a large number of new analytic results concerning chiral dynamics 
with Wilson fermions have been obtained from Wilson chiral perturbation 
theory by working from the perspective 
of the Wilson Dirac eigenvalues \cite{DSV,ADSVNf1,ADSVprd,SVdyn,AN,KVZ,Larsen,SVtwist,KSV,Asger,Mario,Necco:2011vx}. The aim in this review is to give an 
introduction to the new understanding which the results has brought 
about rather than on the technical details.

We will focus on three of the most important new insights which follow from 
the results namely, 
{\sl 1)} that the Sharpe-Singleton scenario is only realized in unquenched 
simulations, {\sl 2)} that the $a/\sqrt{V}$ scaling of the smallest 
eigenvalues of the Hermitian Wilson Dirac operator, observed on the lattice 
\cite{CERN1,CERN2,CERN3}, is a good sign, and finally {\sl 3)} that the results 
offer a new way to measure the additional low energy constants of Wilson 
chiral perturbation theory.

Two new results are also included in this review. {\sl First}, we determine the 
meta-stable region around the Sharpe-Singleton 1st order phase transition, 
and {\sl second} we show that the sampling of the individual sectors is 
essential in order to determine the vacuum structure in the Aoki phase. 

\section{Wilson Chiral Perturbation Theory in the $\epsilon$-regime}

The Wilson term in the Wilson Dirac operator 
\be
D_W =\frac{1}{2}\gamma_\mu(\nabla_\mu+\nabla_\mu^*)
     -\frac{ar}{2}\nabla_\mu\nabla_\mu^*
\ee
breaks chiral symmetry. Therefore, the shortest length scale $a$ also 
affects the physics of longest length scale $1/m_\pi$. This effect 
is captured by Wilson chiral perturbation theory 
\cite{creutz,SharpeSingleton,RS,BRS}, for which we now give a focused review. 

As always in chiral perturbation theory it is essential to set up a 
counting scheme. Here we will work in the $\epsilon$-regime where 
the $V\to\infty$ limit is tied to the chiral and continuum 
limit such that 
\be
 mV, \quad  \zeta V, \quad {\rm and} \quad a^2V 
\ee
are kept fixed and of order unity ($\zeta$ is the axial mass). The partition 
function of Wilson chiral perturbation theory at leading order then reduces 
to a group integral \cite{Shindler,BNSep}
\be
\label{Z}
Z_{N_f}(m,\zeta;a) =   \int_{SU(N_f)} \hspace{-1mm} d U ~e^{S[U]}, 
\ee
where the action is 
\be\label{lfull}
S & = & \frac{m}{2}\Sigma V{\rm Tr}(U+U^\dagger)+
\frac{\zeta}{2}\Sigma V{\rm Tr}(U-U^\dagger)\\
&&-a^2VW_6[{\rm Tr}\left(U+U^\dagger\right)]^2
     -a^2VW_7[{\rm Tr}\left(U-U^\dagger\right)]^2 
-a^2 V W_8{\rm Tr}(U^2+{U^\dagger}^2) .\nn
\ee
The convention for the low energy constants $W_6$, $W_7$ and $W_8$
is that of \cite{DSV,ADSVprd}. In \cite{BRS} these constants are 
denoted respectively by $-{W_6}'$,  $-{W_7}'$ and $-{W_8}'$.

As in the continuum \cite{LS} it is most useful to consider Wilson chiral 
perturbation theory in sectors with fixed index $\nu$, such that 
\cite{DSV,ADSVprd}
\be
\label{Znu}
Z_{N_f}^\nu(m,\zeta;a) =   \int_{U(N_f)} \hspace{-1mm} d U \ {\det}^\nu U
~e^{S[U]}, 
\ee
where we obviously have
\be
\label{Z-Znu}
Z_{N_f}(m,\zeta;a) =\sum_\nu  Z_{N_f}^\nu(m,\zeta;a). 
\ee

The index defined in the above decomposition of the partition function is
the index of the Wilson Dirac operator \cite{DSV,ADSVprd}  
\be
\label{defIndex}
\nu = \sum_{k} {\rm sign} (\langle k|\gamma_5|k\rangle),
\ee
where $|k\rangle$ is the eigenvector associated with the $k$'th real mode
of the Wilson Dirac operator in the physical branch. 
We will make explicit use of this index in section \ref{sec:D5} 
where we discuss the microscopic spectral density. In the first 
sections we will work at mean field level where the dependence on the 
index is suppressed.

From the action in Eq.~(\ref{lfull}) it is clear that the partition function 
in the $\epsilon$-regime of Wilson chiral perturbation theory only depends 
on the dimensionless scaling variables 
\be
 \hm = mV \Sigma,  \qquad
 \hze = \zeta V \Sigma \quad {\rm and}   \qquad \ha_i^2 = a^2 V W_i.
\ee
The mean field limit corresponds to the saddle point approximation to the 
group integral in the limit where these dimensionless 
numbers are all much larger than unity.

\begin{center}
\begin{figure}[t*]
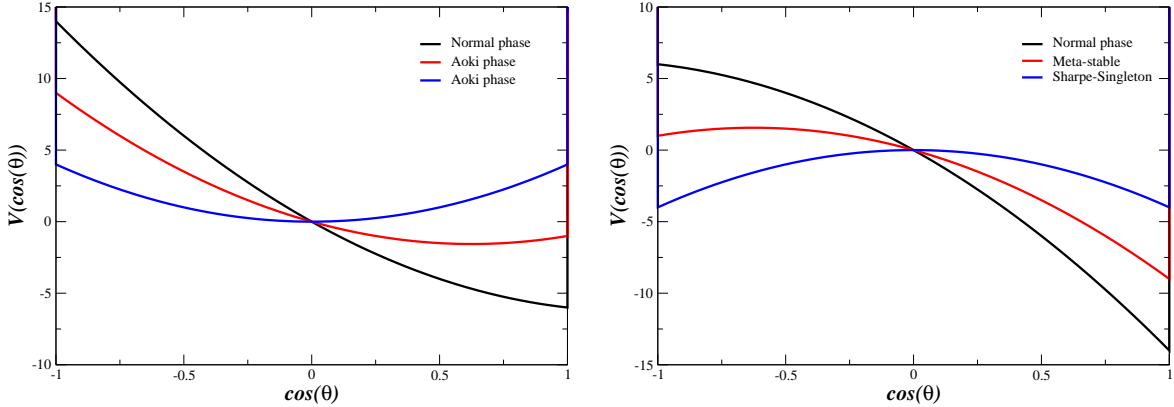
\vspace{2mm}
\centerline{\includegraphics[width=7.5cm]{Aoki-pot-2.eps}\hspace{5mm}\includegraphics[width=7.5cm]{SharpeSingleton-pot-2.eps}}
\caption{\label{fig:V} The mean field potential for the orientation of the vacuum. When $\cos(\theta)=\pm1$ the lattice simulation is in the normal phase with the standard chiral condensate. {\bf Left $W_8+2W_6>0$:} The Aoki phase develops when the minimum of the potential is between -1 and 1. {\bf Right $W_8+2W_6<0$:} For the Sharpe-Singleton scenario the minimum is always at -1 or 1. The additional meta-stable minimum at $\cos(\theta)=-1$ (red curve) becomes degenerate with the minimum at $\cos(\theta)=1$ for $m=0$ (blue curve) resulting in the 1st order Sharpe-Singleton phase transition. The parameters chosen for the plots are $\hm=10,5,0$ and $8(\ha_8^2+2\ha_6^2)=\pm4$.}
\end{figure}
\end{center}

\section{Aoki phase and Sharpe-Singleton scenario}

The form of the order $a^2$ terms in Wilson chiral perturbation theory 
are uniquely determined by the way in which the Wilson term breaks chiral 
symmetry. Each of the three new terms comes with a new low energy constant, 
the value of which is not fixed by the flavor symmetries. The sign of these 
three additional low energy constants, however, is determined by the 
$\gamma_5$-Hermiticity of the Wilson Dirac operator  
\be
\gamma_5D_W\gamma_5=D_W^\dagger.
\ee
Only for 
\be
W_6<0, \qquad W_7<0 \qquad {\rm and} \qquad  W_8>0
\ee 
does the Wilson chiral Lagrangian describe lattice QCD with a 
$\gamma_5$-Hermitian Wislon Dirac operator  
\cite{DSV,ADSVprd,HS1,KSV}. 
As we shall now demonstrate these signs give vital information on how
the Aoki phase and the Sharpe-Singleton scenario are realized. 
To show this let us first work out the phase structure of lattice QCD 
with Wilson fermions at small $m$ and $a$ using Wilson chiral perturbation 
theory at mean field level. With the 
mean field ansatz for the orientation of the Goldstone field 
\cite{SharpeSingleton} (here and below we focus on the case of two mass 
degenerate flavors)
\be
U=\cos(\theta)+i\sin(\theta)\sigma_3
\ee
we find the mean field action 
\be
\label{SMF}
S_{MF} = 2\hm\cos(\theta)-4\ha_8^2(2\cos^2(\theta)-1)-16\ha_6^2\cos^2(\theta).
\ee
This effective potential for $\cos(\theta)$ is plotted in figure 
\ref{fig:V} for the 
two cases $W_8+2W_6>0$ (l.h.s.) and  $W_8+2W_6<0$ (r.h.s.). In the first case the 
orientation, as given by $\cos(\theta)$, continuously moves from 1 to -1
as a function of the quark mass. This is the Aoki phase \cite{Aokiclassic}. 
In the second case,
where $W_8+2W_6<0$, the orientation jumps from 1 to -1 at $m=0$. This is 
the first order Sharpe-Singleton phase transition \cite{SharpeSingleton}.

\begin{center}
\begin{figure}[t*]\vspace{2mm}
\centerline{\includegraphics[width=5cm]{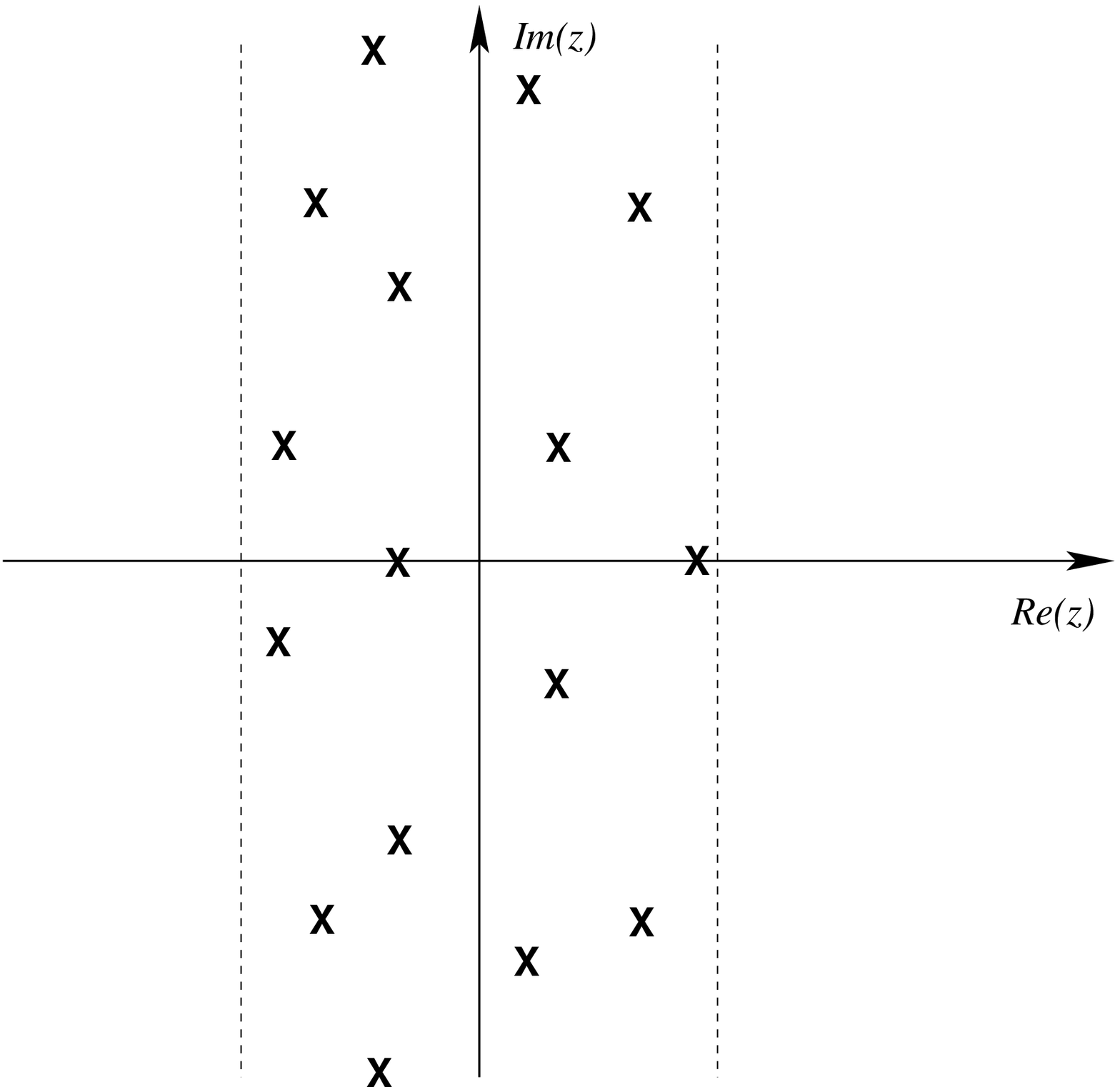}\hfill\includegraphics[width=5cm]{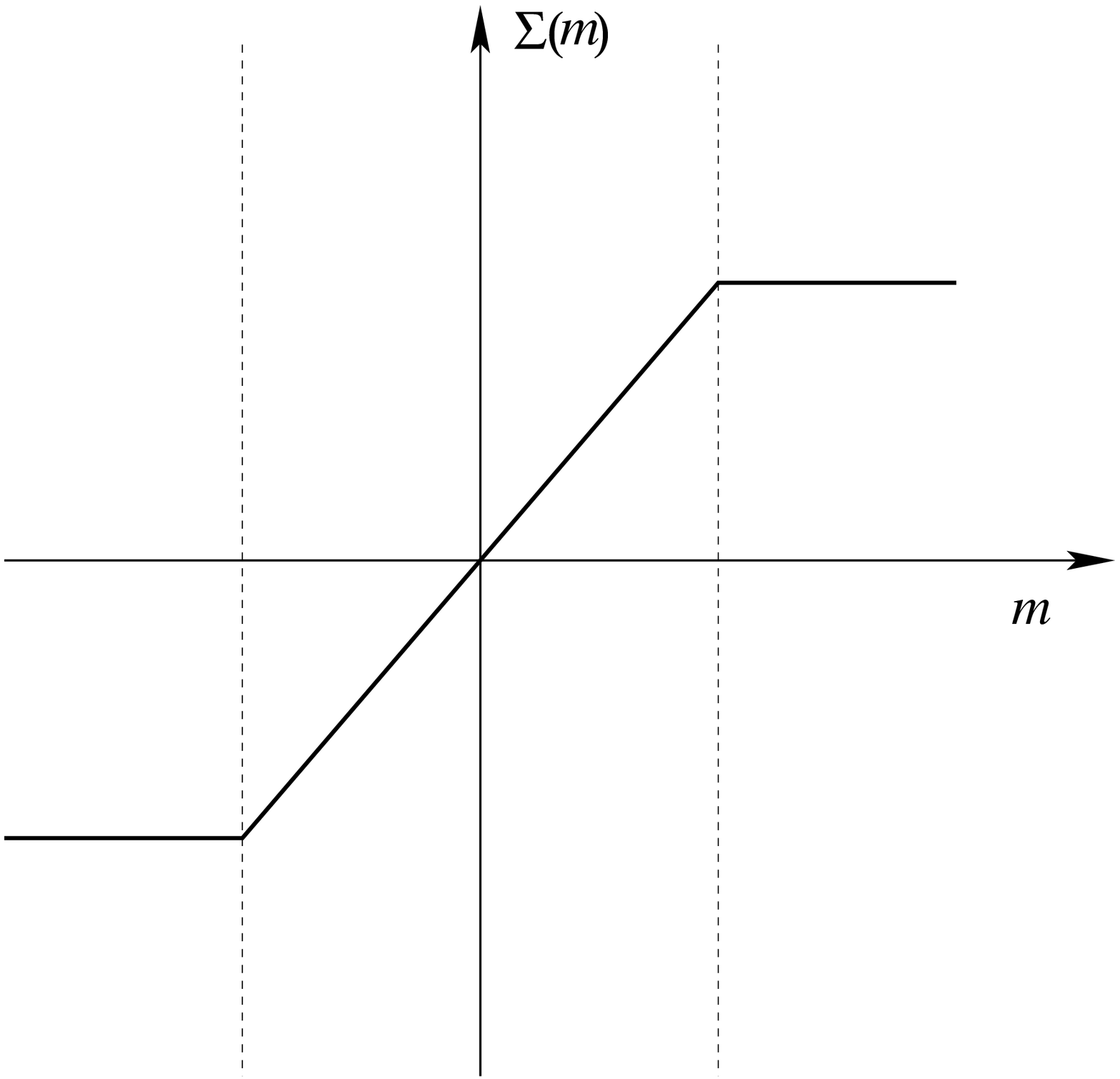}\hfill\includegraphics[width=5cm]{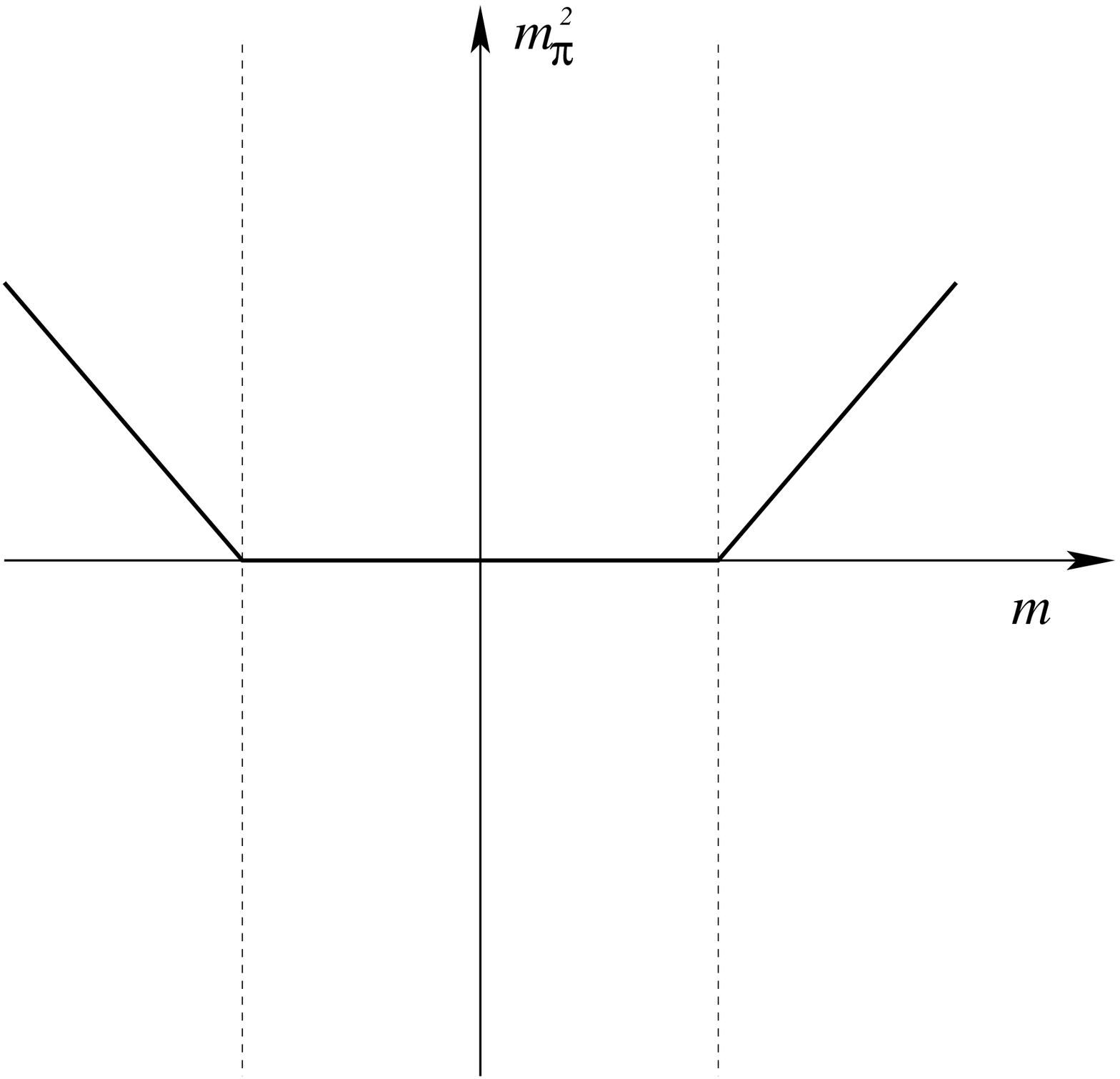}}
\caption{\label{fig:Aoki} The physical branch of the spectral density of the Wilson Dirac operator (left) the chiral condensate (middle) and the pion mass (right). The vertical dashed lines mark the support of the eigenvalue density of the Wilson Dirac operator. The location is at $|Re[z]|\Sigma=8(W_8+2W_6)a^2$. The Aoki phase occours when the quark mass is between the two dashed lines ($W_8+2W_6>0$). In addition there is one massive mode inside the Aoki phase.}
\end{figure}
\end{center}

The pion masses for $|m|\Sigma>8(W_8+2W_6)a^2$ are given by the order-$a^2$ 
corrected GOR-relation \cite{SharpeSingleton,Aoki:masses}
\be
\label{GORaSq}
\frac{m_\pi^2F_\pi^2}{2} = |m|\Sigma-8(W_8+2W_6)a^2.
\ee
Approaching the Aoki-phase (where $W_8+2W_6>0$) the squared mass of 
the pions goes to zero linearly with the quark mass. If we absorb 
the order $a^2$-correction into the quark mass this behavior of the 
pion masses is just like in the continuum. On the contrary in the 
Sharpe-Singleton scenario (where $W_8+2W_6<0$) the pion mass is always 
positive. See the rightmost panels of figure \ref{fig:Aoki}
and figure \ref{fig:realization} respectively.

\begin{figure}[t!]
\vspace{2mm}
\centerline{\includegraphics[width=8cm,angle=0]{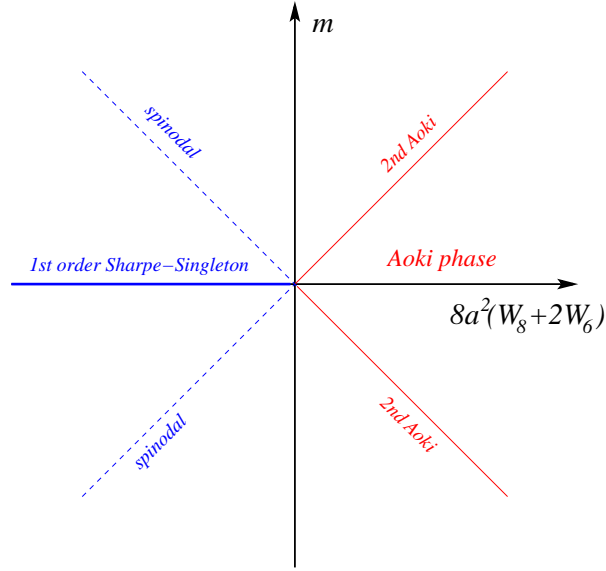}}
\caption{\label{fig:spinodal} The phase diagram for lattice QCD with 
Wilson fermions in the $(8(W_8+2W_6)a^2,m)$-plane. For positive values of 
$8a^2(W_8+2W_6)$ the Aoki phase dominates with the 2nd 
order phase transition at $|m|=8a^2(W_8+2W_6)$ and associated Goldstone 
modes for smaller values of $|m|$. With negative values of $8a^2(W_8+2W_6)$ 
the Sharpe-Singleton scenario dominates with the 1st order phase transition
at $m=0$ and spinodal lines at $|m|=|8a^2(W_8+2W_6)|$. The meta-stable region
is hence between the dashed blue lines.}
\end{figure}

Since the Sharpe-Singleton scenario involves a first order phase transition 
it is natural to ask: What is the meta-stable region? The answer is readily 
seen from Eq.~(\ref{SMF}) or graphically from 
the r.h.s.~of figure \ref{fig:V}. The spinodal line occurs when the {\sl maximum} 
of the effective potential enters between $-1$ and $1$. That is, at exactly the 
same quark mass as the Aoki phase would have occurred for the opposite sign of 
$W_8+2W_6$. This is illustrated in figure \ref{fig:spinodal}.

\section{A puzzle}
\label{sec:puzzle}

The above analysis shows that either the Aoki phase or the Sharpe-Singleton 
scenario will dominate if we take the chiral limit prior to the continuum 
limit. Which of the two is realized, depends on whether $W_8$ is more positive 
than $2W_6$ is negative and hence on the details of the specific lattice 
simulation in question. This analysis was extended to the quenched case in  
\cite{GSS} where it was argued that the Aoki phase and the Sharpe-Singleton 
scenario potentially both should be realizable also in quenched simulations.
On the lattice, however, quenched lattice simulations 
consistently observe the Aoki phase 
\cite{Aoki:1992nb,Aoki:1990ap,Aoki:1989rw,Jansen:2005cg}, while in 
unquenched simulations
both the Aoki and the Sharpe-Singleton scenario are observed \cite{Aoki:1995yf,Aoki:1997fm,Ilgenfritz:2003gw,CERN1,CERN2,CERN3,Ishikawa:2007nn,Borsanyi:2012uq,Aoki:2004iq,BBS,Farchioni:2004us,Farchioni:2004fs,Farchioni:2005tu,Baron}.
As we now show this puzzle has a natural solution when we consider the problem 
from the perspective of the Wilson Dirac eigenvalues.

\section{Spectrum of the Wilson Dirac operator for $W_6=W_7=0$}

We now derive the spectral density of the Wilson Dirac operator. Let us 
start with the case where $W_6=W_7=0$. In this case $W_8+2W_6>0$, 
since $W_8>0$ by the $\gamma_5$-Hermiticity of the Wilson Dirac operator, 
and we have the Aoki phase.

Because of the $\gamma_5$-Hermiticity the eigenvalues of the Wilson Dirac 
operator are either purely real or come in complex conjugate pairs \cite{Itoh}.
For a scatter plot of the eigenvalues, see for example \cite{Anna}. 
In order to derive the spectral density of the Wilson Dirac operator in 
the complex plane 
\be
\rho_{c,N_f}(z,z^*) = \langle \sum_i \delta^2(\lambda^W_i-z) \rangle_{N_f}
\ee
from Wilson chiral perturbation theory we therefore need to extend the 
partition function into a replicated generating functional with $p$ 
additional pairs of conjugate quarks with masses $\hz$ and $\hz^*$,
\be
\label{ZnuNf+2|2}
Z_{N_f+2p}^\nu(\hz,\hz^*,\hm;\ha_i) = \langle \prod_i(\lambda^W_i-m) 
\prod_i(\lambda^W_i-z)^p((\lambda^W_i)^*-z^*)^p \rangle.
\ee
The eigenvalue density of the Wilson Dirac operator in the complex plane 
is then obtained as
\be
\rho_{c,N_f}^\nu(\hz,\hz^*,\hm;\ha_i) 
& = & \partial_{\hz^*}\lim_{p\to0}\partial_{\hz} \frac{1}{p}\log Z_{N_f+2p}^\nu(\hz,\hz^*,\hm;\ha_i).
\ee
The trick is now to use the low energy effective theory, ie. Wilson chiral 
perturbation theory, to compute the replicated generating functional and 
thereby obtain the density. For an introduction to the use of the replica 
method in chiral perturbation theory, see \cite{replica}.

The low energy limit of the replicated generating functional is given by Wilson 
chiral perturbation theory with the mass matrix
\be
\label{replicatedM}
{\cal M}={\rm diag}(\hm_{N_f},\hz_{p},\hz^*_{p}),
\ee
where the subscript on the quark masses refers to the number of times 
the specific mass appears.
In the mean field limit the dependence on the number, $p$, of replica flavors 
is trivial for $W_6=W_7=0$, and one simply obtains (we use the notation $\hz=\hx+i\hy$)
\be
\label{rhoc_MF_a8}
\rho_{c,N_f=2}^{\rm MF}(\hx,\hm;\ha_8)=\theta(8\ha_8^2-|\hx|).
\ee
Note that the eigenvalue density is independent of the imaginary part 
of the eigenvalue, hence it forms a strip along the imaginary axis of width 
$8\ha_8^2$. The Aoki-phase sets in when the quark mass enters the eigenvalue 
strip, as in figure \ref{fig:Aoki}.    

\begin{center}
\begin{figure}[t*]\vspace{2mm}
\includegraphics[width=7cm,angle=0]{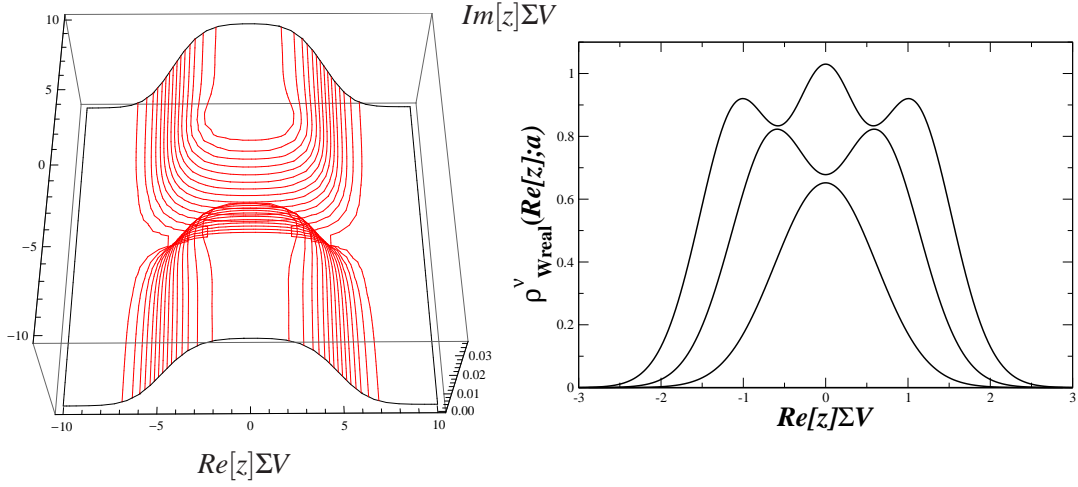}\includegraphics[width=7.5cm]{rho_topo_nu1234_a0p2-v3.eps}\\
\begin{picture}(15,15)
\put(80,0){$Re[z]\Sigma V$}
\put(180,170){$Im[z]\Sigma V$}
\end{picture}
\vspace{2mm}
\caption{\label{fig:DW} {\bf Left:} The microscopic eigenvalue density of the 
Wilson Dirac operator in the complex plane (in the sector with $\nu=0$ for $N_f=0$ and $a\sqrt{W_8V}=0.75$, $W_6=W_7=0$) and {\bf Right:} the 
microscopic eigenvalue density of the real eigenvalues of the Wilson 
Dirac operator (in the sectors with $\nu=1,2,3$ for $N_f=0$ and $a\sqrt{W_8V}=0.2$, $W_6=W_7=0$). The exponential fall-off of the 
eigenvalue density is essential to avoid exceptional configurations. 
}
\end{figure}
\end{center}

The microscopic limit of both the eigenvalue density of $D_W$ in the complex 
plane and on the real axis was worked out in \cite{DSV,ADSVprd,KVZ,KSV}, see 
figure \ref{fig:DW}. The exact way in which the real eigenvalues are 
distributed shows at which value of the quark mass the simulations are safe 
from exceptional configurations. 

\section{The realization of the Sharpe-Singleton scenario: The solution to the puzzle}

\begin{center}
\begin{figure}[t*]\vspace{2mm}
\centerline{\includegraphics[width=5cm]{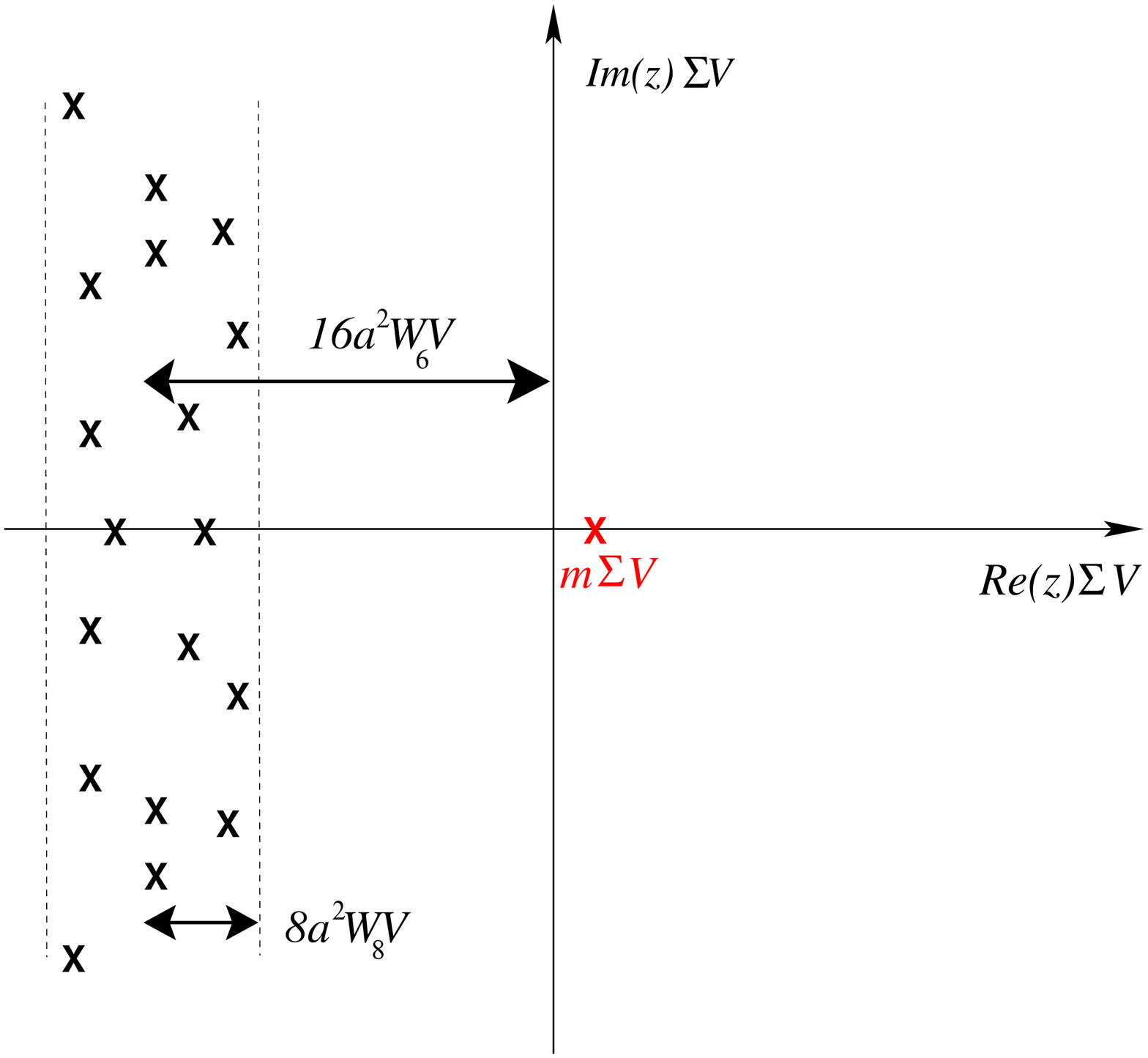}\hfill\includegraphics[width=5cm]{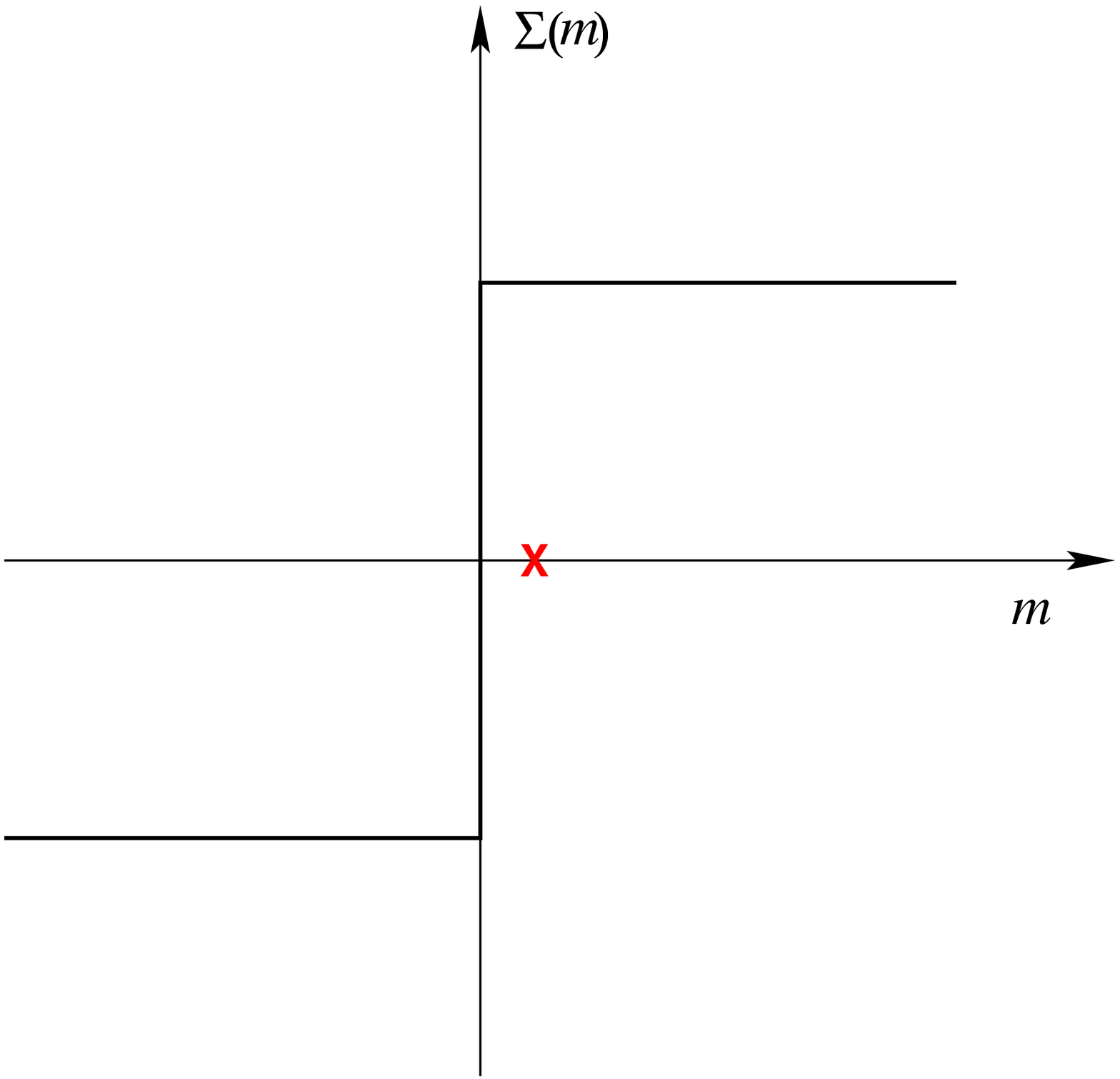}\hfill\includegraphics[width=5cm]{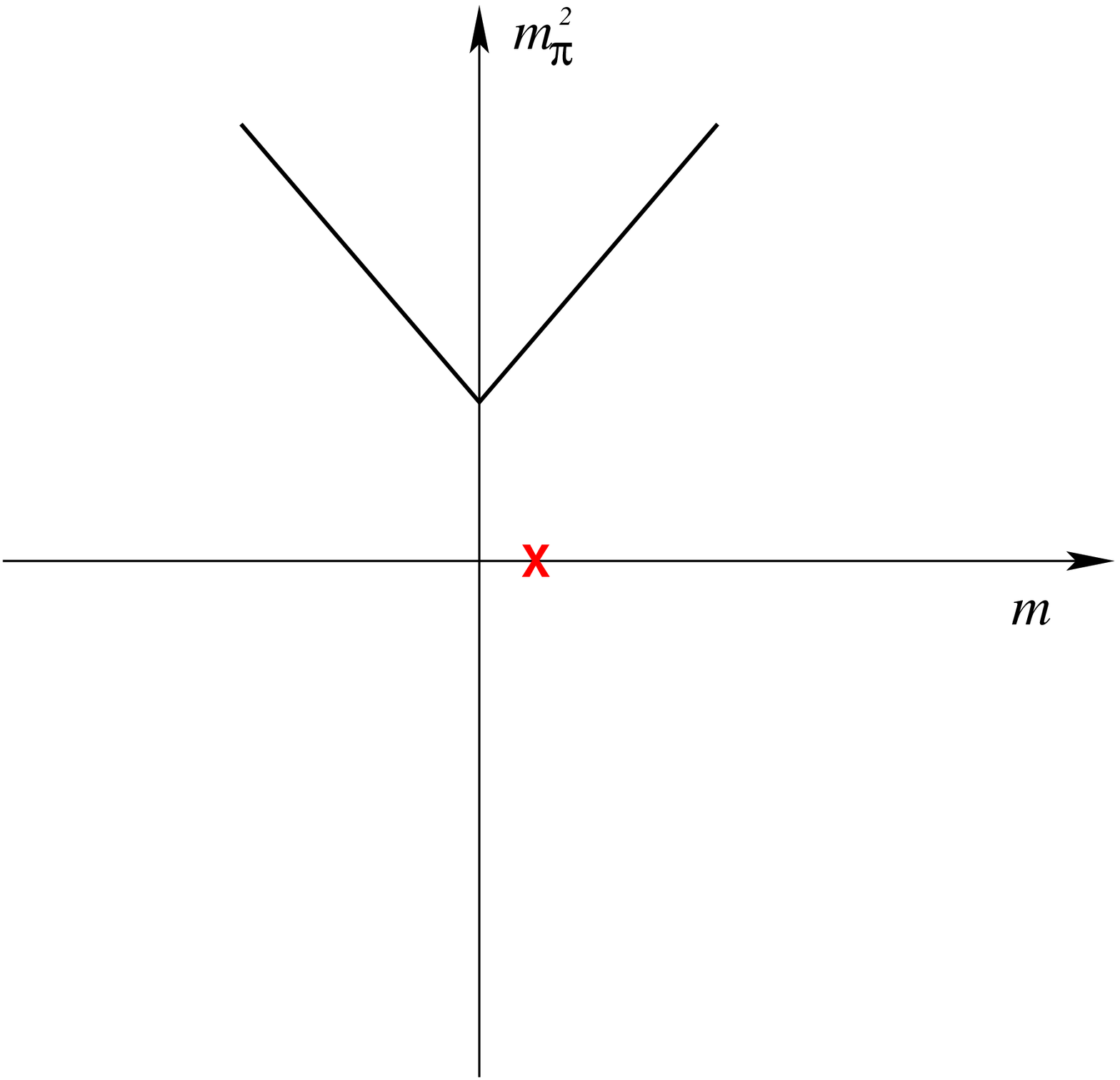}}
\vfill
\centerline{\includegraphics[width=5cm]{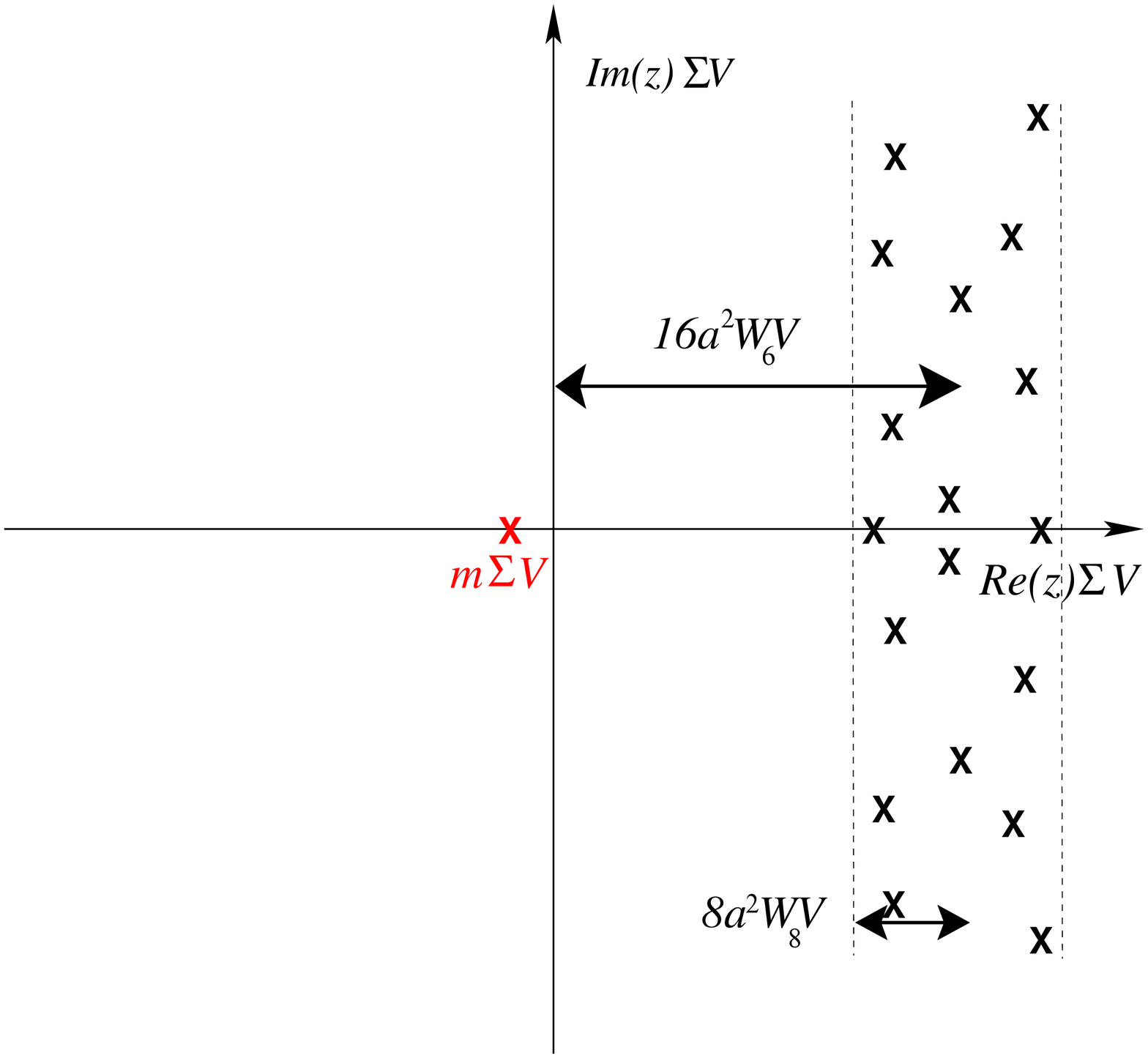}\hfill\includegraphics[width=5cm]{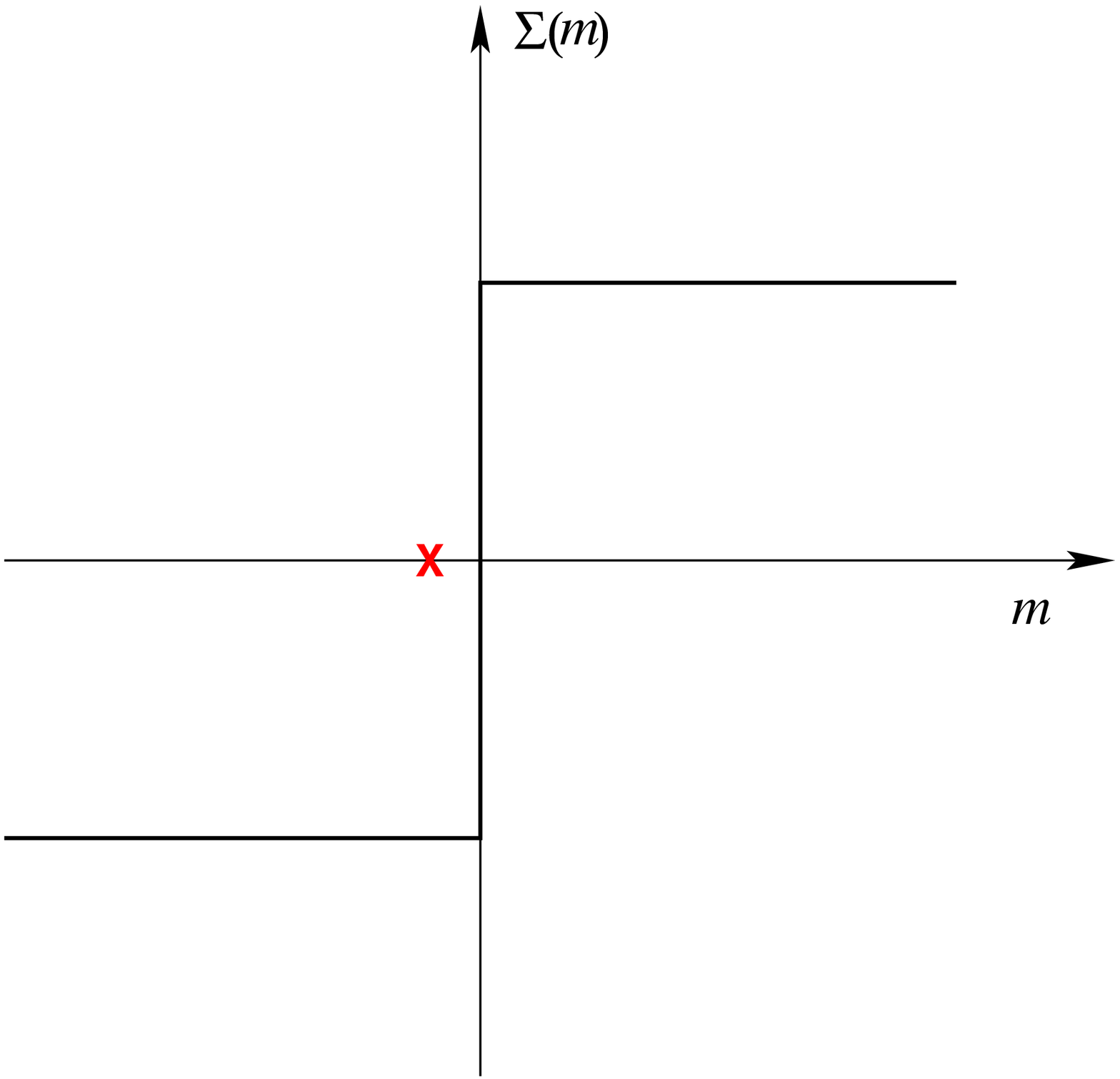}\hfill\includegraphics[width=5cm]{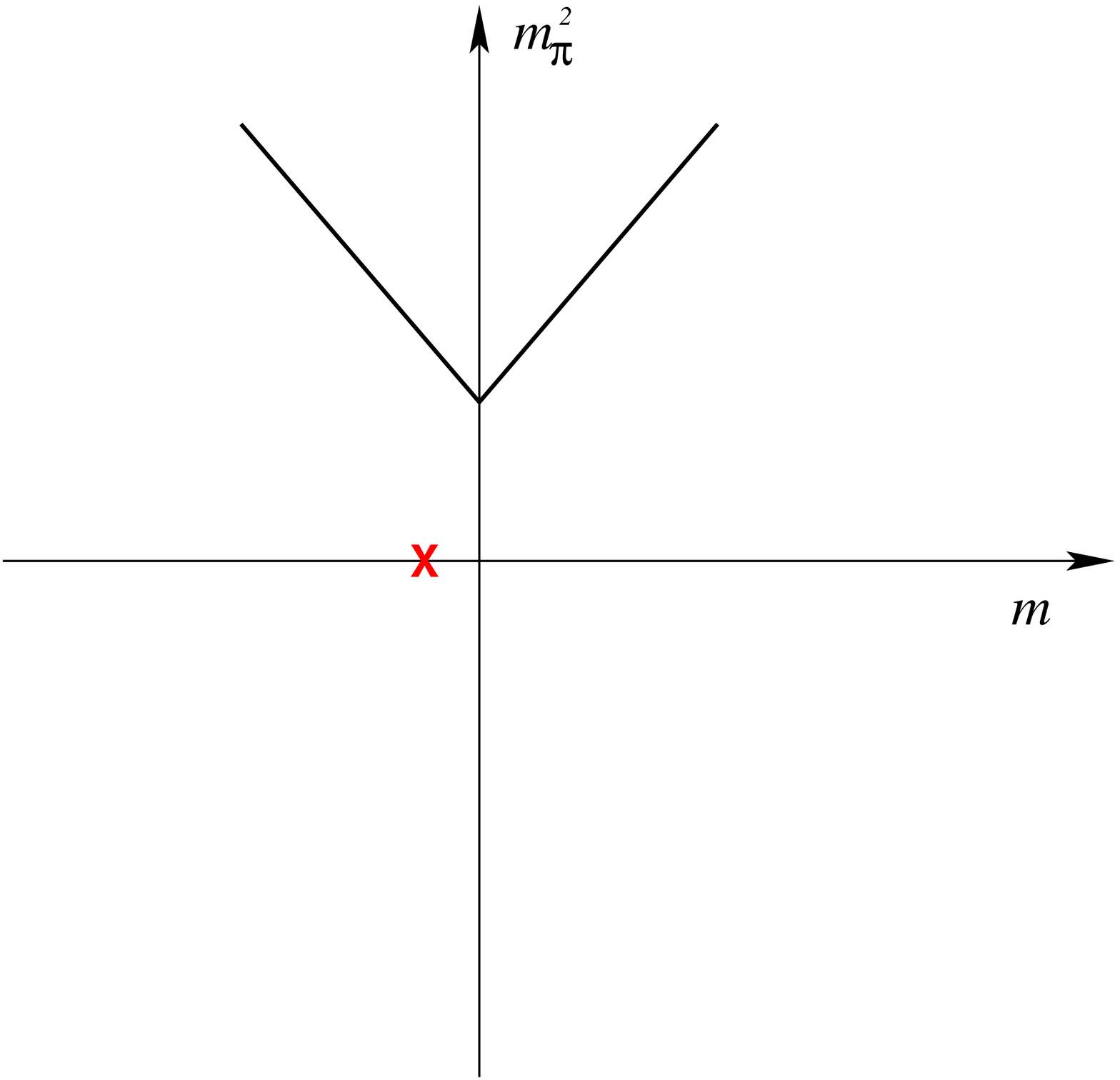}}
\caption{\label{fig:realization} The realization of the Sharpe-Singleton scenario ($W_8+2W_6<0$). The red cross marks the size of the quark mass {\bf top:} the mass is positive and {\bf below:} the mass is negative. As the quark mass changes sign the strip of eigenvalues jumps across the origin and consequently the chiral condensate has a first order discontinuity at $m=0$, cf.~the middle panels. Since the quark mass never reaches the eigenvalue density the pions remain massive even at $m=0$. See the rightmost figures. Clearly this scenario is only possible in the unquenched theory where the eigenvalue density can depend on the quark mass.}
\end{figure}
\end{center}

Let us now compute the effect of $W_6$ on the spectral density of $D_W$ 
in the mean field limit. To do so it is useful first to linearize the 
square in the $W_6$ term of the replicated generating functional with 
a Gaussian integral over a fluctuating mass $y_6$. This leads to \cite{KSV}
\be
\label{rhoc_a6a8}
\rho_{c,N_f}^\nu(\hz,\hz^*,\hm;\ha_6,\ha_8) 
& = & \frac{1}{Z_{N_f}^\nu(\hm;\ha_6,\ha_8)}\int [dy_6] \ Z_{N_f}^\nu(\hm-y_6;\ha_8)\rho_{c,N_f}^\nu(\hz-y_6,\hz^*-y_6,\hm-y_6;\ha_8) , \nn\\
\ee
where we used the notation 
$[dy_6]=dy_6/(4\sqrt{\pi}|\ha_6|)\exp(-y_6^2/(16|\ha_6^2|))$. This expresses 
the eigenvalue density of $D_W$ in the complex plane at $W_6\neq0$ in terms 
of the eigenvalue density with $W_6=0$. 

At the mean field level for $N_f=2$, using Eq.~(\ref{rhoc_MF_a8}), we get
\be
\rho_{c,N_f=2}^{\rm MF}(\hx,\hm;\ha_6,\ha_8)=\frac{1}{Z_2^{\rm MF}(\hm;\ha_6,\ha_8)} 
\int dy_6 \ e^{-y_6^2/16|\ha_6^2|}Z_2^{\rm MF}(\hm-y_6;\ha_8)\theta(8\ha_8^2-|\hx-y_6|),
\ee
where
\be
Z_2^{\rm MF}(\hm;\ha_6,\ha_8) & = & 
e^{2\hm+16|\ha_6^2|-4\ha_8^2}+e^{-2\hm+16|\ha_6^2|-4\ha_8^2} \\
&&+\theta(8(\ha_8^2+2\ha_6^2)-|\hm|)e^{\hm^2/8(\ha_8^2-2|\ha_6^2|)+4\ha_8^2}.
\nn
\ee
This saddle point structure of the two flavor partition function leads to  
\be
\label{rhocNf2MF}
\rho_{c,N_f=2}^{\rm MF}(\hx,\hm;\ha_6,\ha_8) & = & \frac{1}{Z_2^{\rm MF}(\hm;\ha_6,\ha_8)} 
\\ &&
\hspace{-1.3cm}\times \Big\{ e^{2\hm+16|\ha_6^2|-4\ha_8^2}\theta(8\ha_8^2-|\hx+16|\ha_6|^2|)\nn\\ &&
\hspace{-1cm}+e^{-2\hm+16|\ha_6^2|-4\ha_8^2}\theta(8\ha_8^2-|\hx-16|\ha_6|^2|)
\nn\\
&&\hspace{-1cm}+\theta(8(\ha_8^2+2\ha_6^2)-|\hm|)
\theta\left(8\ha_8^2-\left|\hx+\frac{2|\ha_6|^2\hm}{(\ha_8^2-2|\ha_6|^2)}\right|\right)
e^{\hm^2/8(\ha_8^2-2|\ha_6^2|)+4\ha_8^2}\Big\}\nn.
\ee
Since we seek to understand the realization of the Sharpe-Singleton scenario 
let us consider $W_8+2W_6<0$. The density in this case is shown in figure  
\ref{fig:realization}. As the quark mass changes sign the strip of eigenvalues
jump from one side of the origin to the opposite side and this causes the 
1st order jump of the chiral condensate. This explains how the Sharpe-Singleton 
scenario is realized in terms of the strong mass dependence of the 
unquenched eigenvalue density. 

\begin{center}
\begin{figure}[t*]
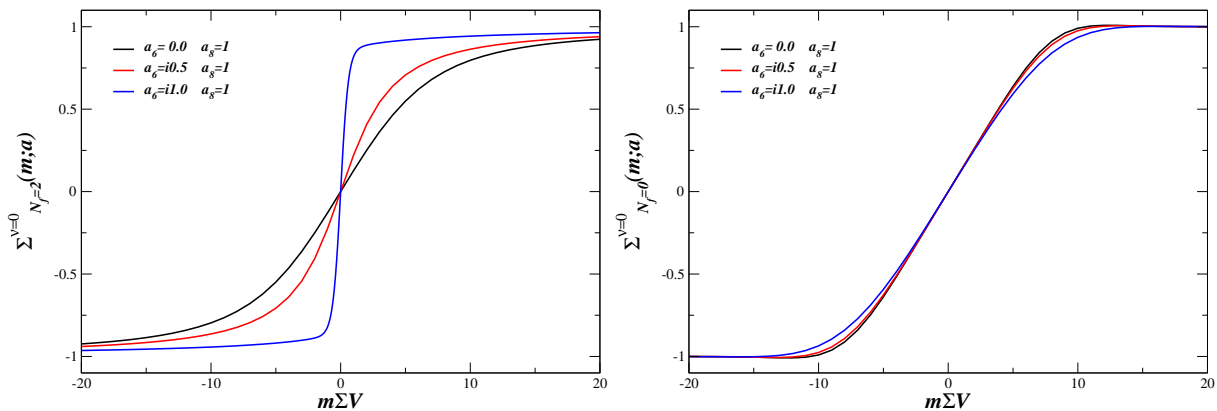
\vspace{2mm}
\includegraphics[width=8cm,angle=0]{Wilson-Nf2-cond-nu0.eps}
\hfill
\includegraphics[width=8cm,angle=0]{Wilson-quenched-cond-nu0.eps}
\caption{\label{fig:cond} The Sharpe-Singleton first order phase 
transition is only realized in the unquenched theory. Here we show 
the mass dependence of the microscopic chiral condensate for 
$\ha_8=1$ and $\ha_6=0,0.5i$ and $i$ corresponding to $W_8+2W_6>0$, 
$W_8+2W_6=0$ and $W_8+2W_6<0$, respectively. {\bf Left} $N_f=2$: 
The 1st order Sharpe-Singleton phase transition develops with the 
increasingly negative value of $W_6$. {\bf Right} $N_f=0$: The smooth 
behavior of the chiral condensate as a function of the quark mass 
characteristic of the Aoki phase is present in the quenched theory 
even when $W_8+2W_6<0$.}
\end{figure}
\end{center}

{\sl In the quenched case} on the contrary the eigenvalue density is 
independent of the quark mass since the fermion determinant 
is absent in the measure. The quark mass in the quenched case, therefore, 
necessarily passes through the eigenvalue density and this leads to the 
standard Aoki phase, {\sl independent of the sign of} $W_8+2W_6$. 
This explains why the Aoki phase is observed consistently in quenched 
simulations and thus solves the puzzle of section \ref{sec:puzzle}.
One can also compute the quenched chiral condensate directly 
from the graded version of the partition function in Wilson chiral 
perturbation theory and reach the same conclusion. This is illustrated in 
figure \ref{fig:cond}. 
The analysis of \cite{GSS} reached the conclusion that 
the Sharpe-Sharpe-Singleton scenario could be realized in 
quenched simulations because the bound on the signs of the $W_i$'s 
due to $\gamma_5$-Hermiticity had not been understood at the time.  

The meta-stable region surrounding the 1st order Sharpe-Singleton phase
transition (see figure \ref{fig:spinodal}) can also be observed in the 
eigenvalue density. For $0<m\Sigma<|W_8+2W_6|a^2$ and $W_8+2W_6<0$ the
meta-stable minimum of $Z_2^{\rm MF}$ introduces a local minimum of
the $y_6$ integral in Eq.~(\ref{rhocNf2MF}). This local minimum offers 
the possibility for the strip of eigenvalues to temporarily jump to the
opposite side of the origin thus inducing a fluctuating sign of the 
chiral condensate. Such fluctuations become more frequent as the 
meta-stable and global minimum of $Z_2^{\rm MF}$ become almost degenerate, ie. as 
$m$ approaches zero. Thus the meta-stability of the eigenvalue density 
is completely consistent with the meta-stability of the chiral condensate 
obtained directly from Eq.~(\ref{SMF}).

It is also instructive to consider the distance from the quark mass to the 
edge of the strip of eigenvalues of $D_W$. From  Eq.~(\ref{rhocNf2MF}) we 
see that this distance is given by
\be
|m|-8(W_8+2W_6)a^2/\Sigma.
\ee
This is exactly the combination which enters the right hand side of the 
order-$a^2$ corrected GOR relation, cf.~Eq.~(\ref{GORaSq}).
The distance from the quark mass to the edge of the strip of
eigenvalues of $D_W$ can therefore be thought of as the effective 
quark mass which enters the standard continuum form of the GOR relation.

In the Sharpe-Singleton scenario, ie. for $W_8+2W_6<0$, the quark mass 
never reaches the strip of eigenvalues of $D_W$. The minimal distance 
between the quark mass and the eigenvalues is given by $-8(W_8+2W_6)a^2$ 
and hence the smallest value which the pion mass can reach is 
\be
m_\pi = \sqrt{-16(W_8+2W_6)a^2/F_\pi^2}.
\ee
This is of course exactly the same minimal value of the pion mass found 
in \cite{SharpeSingleton}. We now understand how this minimum is linked 
to the behavior of the eigenvalue density in the unquenched theory.

\section{Spectrum of the Hermitian Wilson Dirac operator}
\label{sec:D5}

The $\gamma_5$-Hermiticity of the Wilson Dirac operator makes it 
natural to introduce the Hermitian Wilson Dirac operator, $D_5$, 
defined by
\be
D_5 \equiv \gamma_5 (D_W +  m) ~.
\ee
There is a close analogue of the Banks Casher relation \cite{BC} for the 
spectrum of $D_5$: The lattice artifacts induce eigenvalues of $D_5$ with 
magnitude less than $|m|$ and when these build up a density at zero the 
Aoki phase is reached \cite{BHN}. It is therefore essential  
to understand analytically the dependence of the smallest eigenvalues of 
the Hermitian Wilson Dirac operator in order to study chiral 
dynamics in simulations with Wilson fermions.

\begin{figure}[t!]
\vspace{2mm}
\centerline{\includegraphics[width=8cm,angle=0]{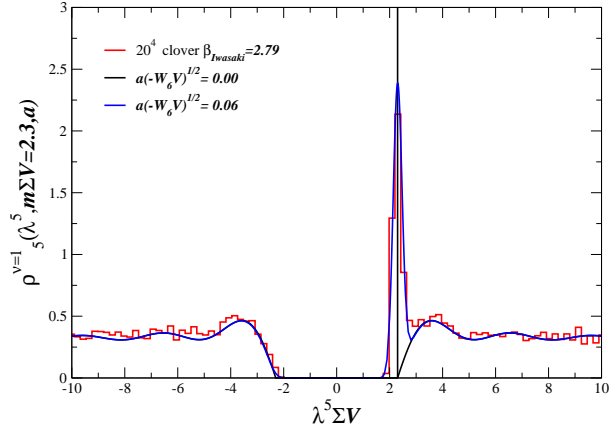}}
\caption{\label{fig:DHS} {\bf Red histogram:} The spectral density 
of the Hermitian Wilson Dirac operator on quenched $20^4$ lattices in the 
sector with $\nu=1$. The eigenvalues have been rescaled
with $\Sigma V/a=220$. See \cite{DHS2} for details about the simulations.
{\bf Blue curve:} The microscopic spectral density obtained from Wilson 
chiral perturbation theory with $N_f=0$, $\nu=1$, $\hm=2.3$ and $\ha_6=0.06$.
Also shown is the {\bf black curve} obtained from ordinary chiral perturbation 
theory with $a=0$. The prime effect of the lattice artifacts for $ha_i\ll1$ 
is to smear the index-peak. This leads to the $a/\sqrt{V}$ scaling of the 
width of the smallest eigenvalue of the Hermitian Wilson Dirac operator 
reported in \cite{CERN1,CERN2}.}
\end{figure}

The first to consider the spectral density of $D_5$ from the 
perspective of Wilson chiral perturbation theory was Sharpe 
\cite{Sharpe1}. By now a detailed analytic understanding of 
the effects the discretization errors have on the spectrum of $D_5$ 
have been obtained \cite{DSV,ADSVprd,SVdyn,AN,KVZ,Larsen,SVtwist,KSV,Asger,Mario,Necco:2011vx}. A plot of the spectral density is shown in figure 
\ref{fig:DHS}. The plot shows the density in the sector with $\nu=1$ 
and the index peak at $m$ is clearly visible. Also the precise way 
the eigenvalues intrude into the region between $-m$ and $m$ can be 
observed. 
 
The predictions also offer a new way to measure the low energy constants 
of Wilson chiral perturbation theory \cite{Necco:2011vx,DHS1,DWW,DHS2}: The 
histogram in figure \ref{fig:DHS} displays lattice data and the match of the 
analytic prediction to the lattice data fixes the values of the low energy 
constants. We refer to \cite{DHS2} for details. 

One can also obtain the low energy constants from a fit of the analytic 
predictions for the spectral density of $D_W$ 
to lattice data. However, since $D_5$ is Hermitian it is somewhat easier 
to determine this spectrum numerically in lattice simulations. See also 
\cite{HS2} for an alternative way to measure the $W_i$'s.

\begin{figure}[t!]
\centerline{\includegraphics[width=15cm,angle=0]{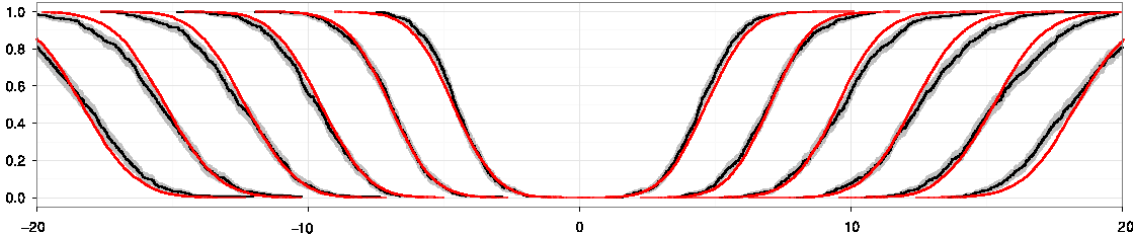}}
\caption{\label{fig:dataBern} The accumulated single eigenvalue distributions 
of the lowest eigenvalues of the Hermitian Wilson Dirac operator. Plot adopted 
from \cite{DWW}.}
\end{figure}

\begin{figure}[t!]
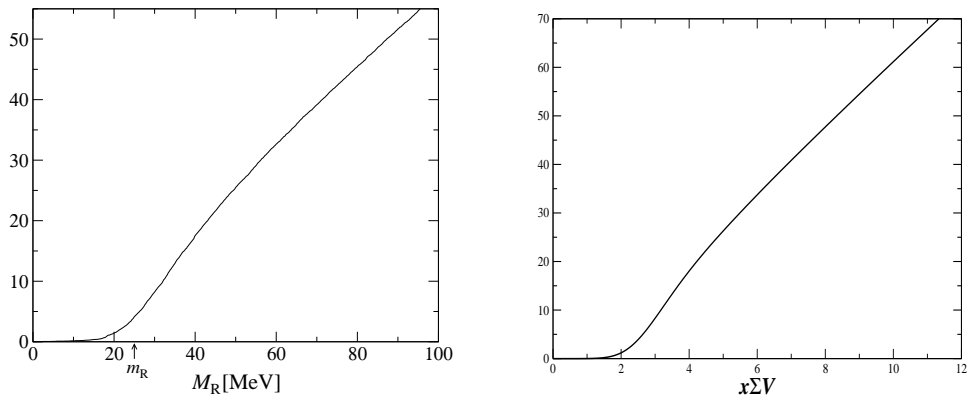

\vspace{2mm}
\centerline{\includegraphics[width=6.1cm]{LuscherPalombi.eps}\hspace{1cm}
\includegraphics[width=5.74cm,height=5.1cm]{rho5sumnuint.eps}}
\caption{
\label{fig:dataCERN} The accumulated eigenvalue density 
the Hermitian Wilson Dirac operator including all sectors. {\bf Left:} 
quenched lattice simulation from \cite{LP}. {\bf Right:} example of the 
analytic prediction from Wilson chiral perturbation theory in the $\epsilon$ 
regime with $N_f=0$. The parameters chosen for this example 
are $m\Sigma V=3$, $\ha_8=0.2$ and $W_6=W_7=0$.}
\end{figure}

\section{Scaling with $a/\sqrt{V}$}

In the simulations of \cite{CERN1,CERN2,CERN3} it was found that 
the width of the 
distribution of the smallest eigenvalue of $D_5$ scales with $a/\sqrt{V}$. 
At first sight this may sound disturbing, since the Banks-Casher
relations \cite{BC,BHN} suggests that the smallest eigenvalues should scale 
with $1/V$. Hence one might fear that the $a/\sqrt{V}$-scaling is a sign of 
uncontrolled lattice artifacts, such as local defects not described by 
Wilson chiral perturbation theory. On the contrary, {\sl the $a/\sqrt{V}$ 
scaling is a good sign}: In the limit where $\ha_i\ll1$ the smallest 
eigenvalues of $D_5$ should in fact scale precisely with $a/\sqrt{V}$ 
\cite{DSV,ADSVprd}. For $\ha_i\ll1$
the primary influence of the lattice artifacts is to smear the topological 
mode into a peak around $m$ of width $a/\sqrt{V}$, as shown in figure \ref{fig:DHS}. 
This fact, obtained from Wilson chiral perturbation theory, has been 
demonstrated explicitly in lattice quenched simulations \cite{DHS1,DWW,DHS2}.
See figures \ref{fig:DHS} and \ref{fig:dataBern}. Note that these lattice 
simulations were separated into sectors with fixed index. The effect 
of the discretization error on the spectrum of $D_5$ can also be studied 
without the separation into sectors. In figure \ref{fig:dataCERN} the 
accumulated eigenvalue density including all sectors is shown. The left 
hand plot is lattice data obtained in \cite{LP} while the right hand plot 
gives an example of the analytic results. It would be most interesting to 
preform a systematic fit to the data including the leading order corrections 
to the slope at large $\lambda^5$ obtained in \cite{Necco:2011vx}.

\section{Alternative vacuum in the Aoki phase for $N_f=2$}

In \cite{Azcoiti} it was proposed that the additional condensate 
$\langle i\bar{u}\gamma_5u+ i\bar{d}\gamma_5d \rangle$
will take a nonzero expectation value inside the Aoki phase
along with the standard Aoki condensate  
$\langle i\bar{u}\gamma_5u- i\bar{d}\gamma_5d \rangle$.
This suggestion is particularly interesting since the presence of the 
additional condensate is in direct contradiction with the 
results of Wilson chiral perturbation theory \cite{SharpeAzcoiti}.
The additional condensate 
$\langle i\bar{u}\gamma_5u+ i\bar{d}\gamma_5d \rangle$ is zero
in the $\epsilon$-regime of Wilson chiral perturbation theory, 
Eq.~(\ref{Z}), because it is the v.e.v.~of
${\rm Tr}(U-U^\dagger)$ which vanishes in $SU(2)$ where 
${\rm Tr}U={\rm Tr}U^\dagger$. 
With fixed index, however, the group average is extended to $U(2)$ 
and ${\rm Tr}(U-U^\dagger)$ does 
not vanish trivially. The behavior of 
$\langle i\bar{u}\gamma_5u+ i\bar{d}\gamma_5d \rangle$ at fixed index 
can be examined by explicit evaluation
of $U(2)$ integral ~Eq.~(\ref{Znu}) and differentiation w.r.t.~$\zeta$. 
One can also evaluate the squared condensate 
$\langle(i\bar{u}\gamma_5u+i\bar{d}\gamma_5d)^2\rangle$ at zero 
external sources as suggested in \cite{AzcoitiNew}. Again this is 
zero in SU(2) and nonzero at fixed index. 
As should of course be true, one recovers the vanishing value of 
the additional condensates within Wilson 
chiral perturbation theory, after a careful summation 
over all sectors with the appropriate partition functions as weights. 
In simulations, however, it can be very tricky to control this sum over 
the sectors. Even a small error in 
the sampling of the sectors can lead to a substantial error in the 
value of the additional condensates after 
summation over the sectors. The summation over the sectors 
become particularly delicate in the Aoki region if the low energy 
constant $W_7$ is non-vanishing. The reason is that the term
proportional to $W_7$ is tantamount to a Gaussian fluctuating source for 
$i\bar{u}\gamma_5u+i\bar{d}\gamma_5d$. In other words, Wilson chiral 
perturbation theory predicts that it is essential to 
sample all the sectors with the correct weight in order to 
determine if the additional condensate is present.

\section{Conclusion}

The approach to chiral dynamics with Wilsons fermions from the 
perspective of the eigenvalues of the Wilson Dirac operator has lead 
to several new insights. On the macroscopic level it 
becomes clear that the Sharpe-Singleton scenario can only be realized 
in unquenched theories and on the microscopic scale we learn that the 
$a/\sqrt{V}$ scaling of the smallest eigenvalues of the Hermitian 
Wilson Dirac operator is a sign that the simulation is very close 
to the continuum. On a more technical but equally important level we
now understand that the signs of the additional low energy constants in 
Wilson chiral perturbation theory are fixed. Moreover, the new analytic 
results offers 
a practical way to measure the additional low energy constants in lattice 
simulations. These combined new insights can be most useful when we 
seek to minimize discretization errors in lattice simulations with 
Wilson fermions. For example, to minimize $W_8+2W_6$   
it is useful to know that the two contributions have opposite signs.

A control of the lattice artifacts becomes particularly important 
in studies of QCD with a larger number of flavors: In a mean field 
approach the coefficient in front of $W_6$ and $W_7$ scales with 
$N_f^2$ compared to the linear scaling of the $W_8$ 
term. This suggest that simulations with a larger number of flavors should 
be more likely to end up in the Sharpe-Singleton scenario. Hence one is less 
likely to be able to exploit the massless modes at the boundary of the 
Aoki phase to mimic the chiral limit for larger number of flavors.

Many aspects discussed in this review of chiral dynamics with Wilson 
fermions has a direct analogue for staggered fermions. It would be 
interesting to pursue this further. The necessary tools are already 
available \cite{JO} and first results have appeared \cite{KVZlat}. 
The spectral properties of the Dirac operator could also cast additional 
light on simulation with mixed actions, see eg.~\cite{NNPV}.  

\vspace{2mm}

\noindent
{\bf Acknowledgments:} I would like to thank the organizers 
of Lattice 2012 for an excellent conference and the chance to 
review these results in the 
plenary session. Furthermore, it is a great pleasure to thank 
Poul Henrik Damgaard, Jac Verbaarschot, Gernot Akemann, Urs Heller and 
Mario Kieburg for the collaborations leading to the publications this 
review is based upon. Joyce Myers is thanked for useful comments on the 
manuscript.  
This work was supported by the {\sl Sapere Aude} program of
The Danish Council for Independent Research.

\end{document}